\documentclass[11pt]{article}

\usepackage{jheppub}
\usepackage[utf8]{inputenc}
\usepackage{comment}

\usepackage{graphicx,ulem}
\usepackage{caption,epigraph}
\usepackage{subcaption}
\usepackage{hyperref}
\usepackage[T1]{fontenc}
\usepackage[export]{adjustbox}
\usepackage[table]{xcolor}
\usepackage{amssymb}
\usepackage{pifont}

\newcommand{\be}{\begin{equation}}
	\newcommand{\ee}{\end{equation}}
\newcommand{\bea}{\begin{eqnarray}}
	\newcommand{\eea}{\end{eqnarray}}

\usepackage{multirow, graphicx,amssymb,url,mathrsfs,amsmath}
\usepackage{amsxtra,amstext,latexsym,dsfont,amsfonts}
\usepackage{color,eucal}
\usepackage{float}
\usepackage{colortbl}
\usepackage{multirow}
\usepackage{tikz}
\usetikzlibrary{arrows}
\usepackage{tabularx, array}
\newcolumntype{L}[1]{>{\raggedright\arraybackslash}p{#1}}
\newcolumntype{C}[1]{>{\centering\arraybackslash}p{#1}}
\newcolumntype{R}[1]{>{\raggedleft\arraybackslash}p{#1}}
\makeatother
\makeatletter
\newcommand{\myBig}{\bBigg@{1.75}}
\makeatother

\title{\Large Dynamical and thermodynamic crossovers in the supercritical region of a holographic superfluid model}

\author[a,b]{Zi-Qiang Zhao,}
\author[a,1]{Zhang-Yu Nie,\note{Corresponding authors.}}
\author[b,1]{Jing-Fei Zhang,}
\author[b,1]{Xin Zhang,}
\author[c,1]{Matteo Baggioli}

\affiliation[a]{Center for Gravitation and Astrophysics, Kunming University of Science and Technology, Kunming 650500, China}
\affiliation[b]{Key Laboratory of Cosmology and Astrophysics (Liaoning) \& College of Sciences, Northeastern University, Shenyang 110819, China}
\affiliation[c]{Wilczek Quantum Center, School of Physics and Astronomy, Shanghai Jiao Tong University \& Shanghai Research Center for Quantum Sciences, Shanghai 200240, China}

\emailAdd{zhaoziqiang@stu.kust.edu.cn}
\emailAdd{niezy@kust.edu.cn}
\emailAdd{jfzhang@mail.neu.edu.cn}
\emailAdd{zhangxin@mail.neu.edu.cn}
\emailAdd{b.matteo@sjtu.edu.cn}

\abstract{Many physical systems, including classical fluids, present in their phase diagram the competition between two phases that are separated by a line of first-order phase transitions which terminates at a so-called critical point. Despite several proposals, in the supercritical region beyond the critical point, whether the two phases can still be distinguished and by which criterion remain open questions. In this work, we study the thermodynamics and linear dynamics of a holographic superfluid model with nonlinear potential terms in the supercritical region. We identify the presence of a dynamical crossover, akin to the liquid-like to gas-like Frenkel transition in supercritical fluids, and we define other separation lines of thermodynamic origin based on higher order derivatives of the free energy with respect to the charge density. Our results highlight the universal dynamical and thermodynamic features of supercritical systems from nuclear matter and classical fluids to superfluid systems.
}

\begin{document}
	\maketitle
	\flushbottom
	\section{Introduction}
\begin{figure}
    \centering
    \includegraphics[width=\linewidth]{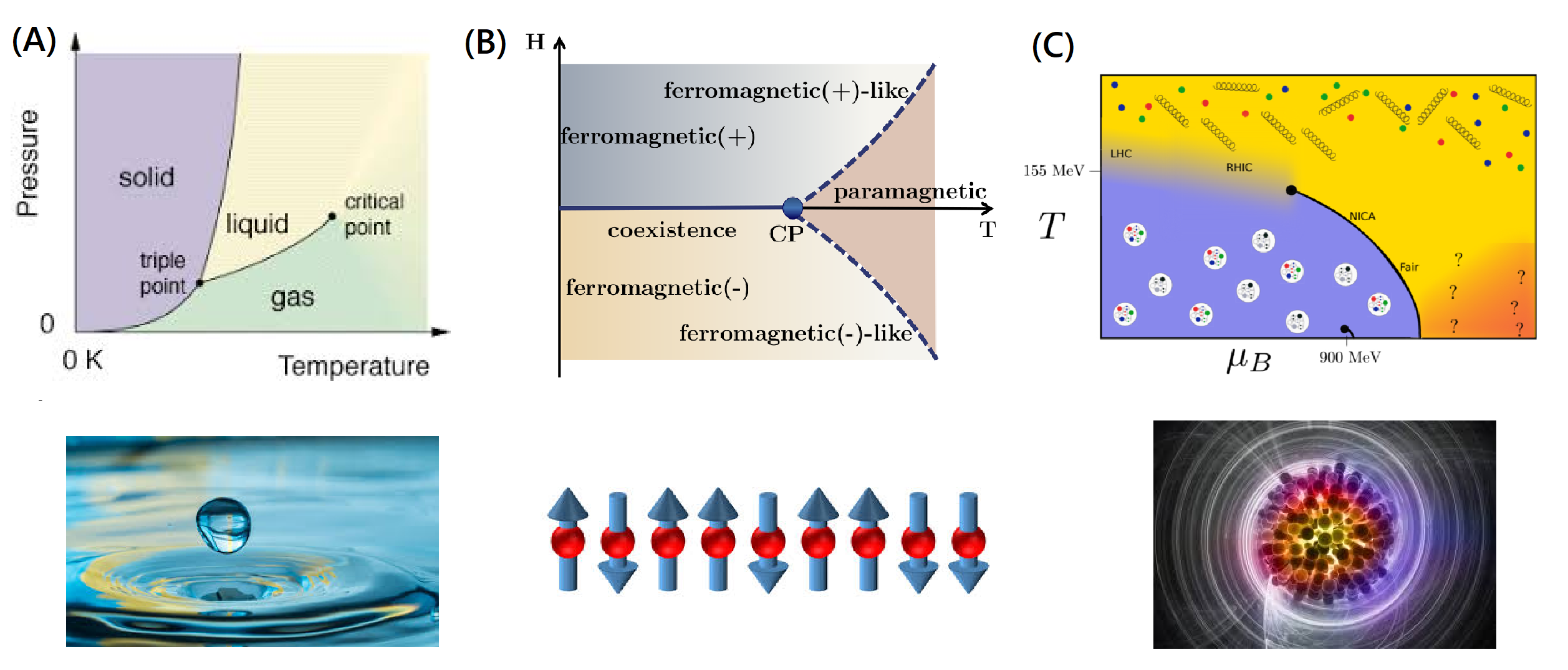}
    \caption{Several physical systems display a phase diagram with a supercritical region emerging beyond a critical point. \textbf{(A)} Classical fluids: the liquid and gas phase are separated by a line of first-order phase transitions that ends at a critical point. \textbf{(B)} Ising model: two different separation lines can be drawn \cite{li2023thermodynamic}. \textbf{(C)} QCD: there exists a supercritical region between a hadron gas phase and a quark-gluon plasma liquid-like phase separated by a continuous crossover below the QCD critical point \cite{PhysRevLett.75.1040}.}
    \label{fig:0}
\end{figure}

In many physical systems (see Figure~\ref{fig:0}), two phases are separated by a first order phase transition that usually ends at a critical point. Beyond the critical point lies the supercritical region, in which the two phases are indistinguishable, \textit{i.e.}, not anymore separated by a sharp thermodynamic transition. The most notable example in this class is that of classical fluids where the liquid and gas phases are not anymore clearly separated for large temperature/pressure \cite{10.1063/PT.3.1796}, \textit{i.e.}, in the supercritical fluid region. In recent years, several different criteria to define additional separation lines in the supercritical region have been proposed. Some of them are of thermodynamic nature, \textit{e.g.}, Widom line \cite{Xu_2005,Ruppeiner_2012,PhysRevLett.112.135701,PhysRevE.95.052120,Gallo2014}; some of structural nature, \textit{e.g.}, Fisher-Widom line \cite{10.1063/1.1671624,RJFLeotedeCarvalho_1994,PhysRevE.51.3146}; and finally, some are dynamical, \textit{e.g.}, Frenkel line \cite{Yoon_2018,PhysRevLett.111.145901,Bolmatov2013,Prescher_2017,Bolmatov_2015,Fomin_2018,Fomin2015,PhysRevE.85.031203,2023PhRvR...5a3149H,jiang2024experimental}. Despite the long history of this debate, a final verdict is still missing and new surprises still emerge nowadays (see \textit{e.g.} Ref.~\cite{li2023thermodynamic}).

Moreover, supercritical systems give rise to many interesting physical problems. For example, the steam in supercritical fluids is commonly used to be the working substance in heat engines to improve their efficiency. However, an underlying theoretical explanation of the enhancement of the efficiency is still lacking since it involves complex non-equilibrium physical processes in a region where many mysteries remain. Inspired by the many questions related to supercritical phases of matter, the lack of controlled theoretical frameworks and their fundamental usage in several chemical and industrial processes, we leverage on the AdS/CFT correspondence to investigate this problem from a different angle.

The AdS/CFT correspondence provides a theoretical tool to deal with strongly coupled quantum systems \cite{hartnoll2018holographic}. Importantly, this framework allows for a description of the physics at all scales, including the far-from-equilibrium regime where complex time dependent dynamics emerge \cite{Liu_2020}. First order phase transitions are considered in AdS/CFT in the context of QCD as well as condensed matter problems. In a recent study, first order superfluid phase transitions are realized in a nonlinearly extended version of the original holographic superfluid model~\cite{Hartnoll:2008vx} in the probe limit, and the dynamical stability as well as the full dynamical evolution are studied in detail~\cite{Zhao:2022jvs,Zhao:2023gcv}. first order phase transitions between two superfluid phases are also realized \cite{Zhao:2022jvs} and give rise to a critical point located at the end of the curve of first order phase transition points, and to a corresponding supercritical region. Therefore, the model studied in Ref.~\cite{Zhao:2022jvs} is convenient for studying the dynamical properties in the supercritical region and searching for separation lines within the latter, similar in spirit to the Frenkel and Widom lines mentioned above.

In this work, we perform a detailed investigation of the thermodynamic and dynamical properties of the supercritical phase of such holographic superfluid model. In Section \ref{sec2}, we present the holographic setup and the phase diagram containing the critical point as well as the supercritical region. In Section \ref{dynsec}, we study the existence of dynamical separation lines within the supercritical phase. In Section \ref{sec4}, we proposed a revised phase diagram for this holographic model taking into account all the dynamical criteria. In Section \ref{termsec}, we analyze possible thermodynamic features in the supercritical region. Finally, we provide some conclusions in Section \ref{sec5}. 
\section{Holographic setup and phase diagram}\label{sec2}
We consider a holographic s-wave superfluid model with two nonlinear self-interaction terms introduced in Ref.~\cite{Zhao:2022jvs} and we work within the probe limit. The background is given by an asymptotically AdS$_4$ black brane defined by the following metric:
\begin{align}
ds^{2}=-f(r)dt^{2}+\frac{1}{f(r)}dr^{2}+r^{2}dx^{2}+r^{2}dy^{2},
\end{align}
where the function $f(r)$ is given by
\begin{align}
f(r)=r^2\left(1-\frac{r_h^3}{r^3}\right)~,
\end{align}
and it is kept fixed in our computation. We choose coordinates such that the AdS boundary is located at $r=\infty$. On top of this geometrical background, we consider the dynamics of a complex scalar field charged under U(1) symmetry and a bulk gauge field $A_\mu$ described by the following action
\begin{align}
&S_M=\int d^{4}x\sqrt{-g}\left(-\frac{1}{4}F_{\mu\nu}F^{\mu\nu}
-D_{\mu}\psi^{\ast}D^{\mu}\psi  -m^{2}\psi^{\ast}\psi-\lambda(\psi^{\ast}\psi)^{2}-\tau(\psi^{\ast}\psi)^{3}\right)~.\label{Lagm}
\end{align}
Here, $D_{\mu}\psi=\nabla_{\mu}\psi-i q A_\mu\psi$ is the standard covariant derivative term and $F_{\mu\nu}=\nabla_{\mu}A_{\nu}-\nabla_{\nu}A_{\mu}$ is the Maxwell field strength. 
In this work, we consider the simple ansatz $\psi=\psi(r)$ and $A_\mu dx^\mu=\phi(r)dt~$. We fix standard quantization for all the bulk fields, so that the chemical potential $\mu$ as well as the charge density $\rho$ are obtained from the boundary behavior of the electric potential using the asymptotic expansion $\phi(r)\approx \mu-\frac{\rho}{r}$. Finally, the complex scalar field presents the following behavior at the AdS boundary,
\begin{align}
    \psi(r)\approx \psi_1 r^{-1} + \langle O \rangle r^{-2}~,
\end{align}
where $\psi_1$ and $\langle O \rangle$ are respectively the source and the expectation value of the dual scalar operator. A solution with $\psi_1=0$ and $\langle O \rangle \neq 0$ corresponds to a superfluid state in the dual field theory that spontaneously break the global U(1) symmetry at the boundary.

We work in the canonical ensemble. We further set the values of the model parameters to be $m^2=-2$, $q=r_h=L=1$, $\lambda=-0.05$ and consider $\tau$ and $\rho$ as free parameters. Before continuing, we note that because of scale invariance and the probe limit approximation, the only sensitive parameter is $\rho/T^2$ and we choose $\rho$ as one thermodynamic parameter in the phase diagram. We also notice that $\tau$ is not strictly speaking a thermodynamic parameter of the boundary field theory. It is rather a coupling of the bulk model, meaning that different values of $\tau$ correspond to different systems at the boundary. Interestingly, couplings of this sort can be thought as thermodynamic parameters in a generalized thermodynamic framework \cite{hajian2023coupling}.
Therefore, being aware of these caveats, we will consider the phase diagram of the model in the generalized $\tau$-$\rho$ plane.
\begin{figure}
\centering
\includegraphics[width=0.7\columnwidth]{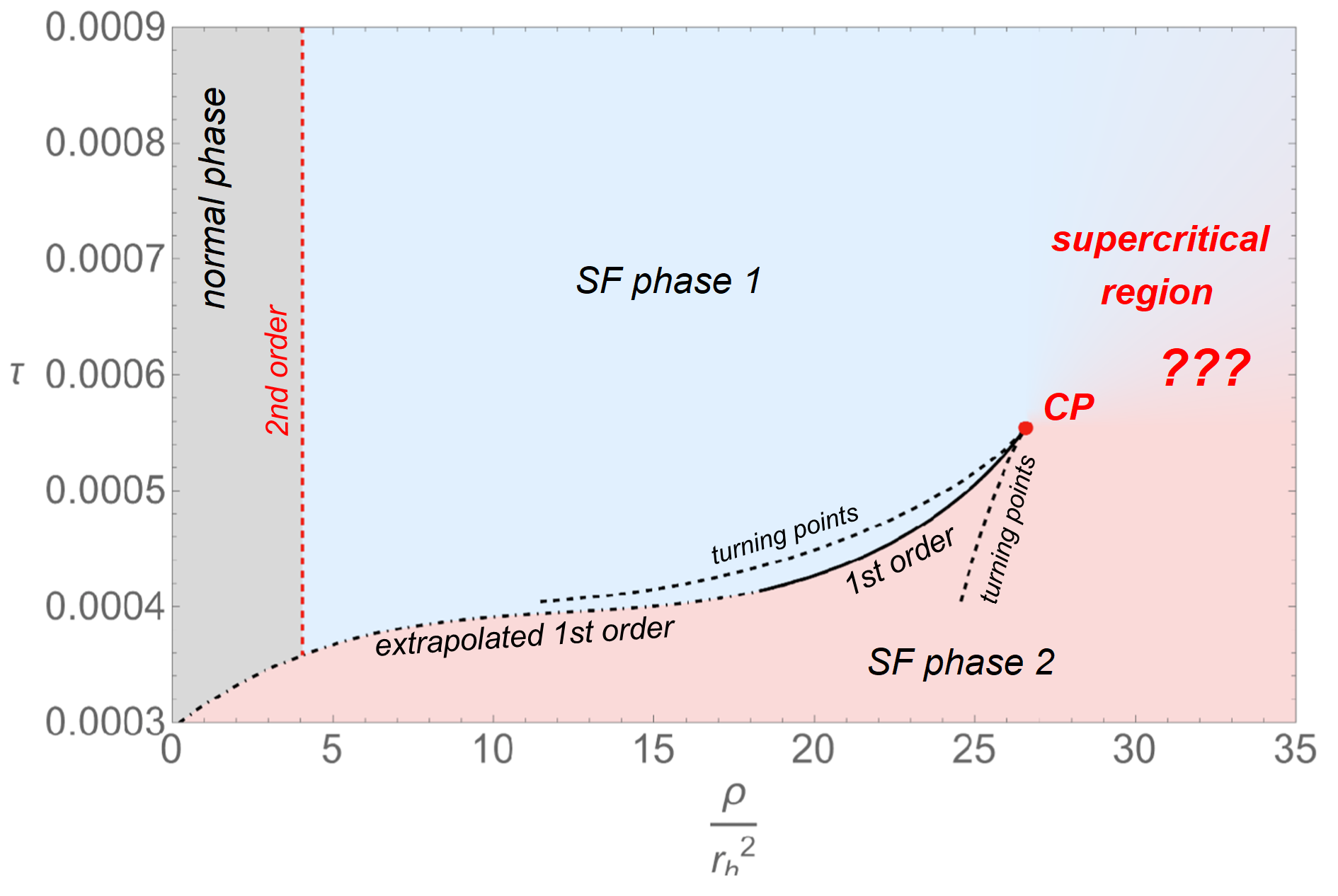}
\caption{The $\tau$-$\rho$ phase diagram with $\lambda=-0.05$. The red dot is the critical point (CP). The red dashed line is a line of second order superfluid phase transitions from the normal phase (gray area) to a superfluid phase with small condensate (light blue area). The solid black line is a line of  first order phase transitions between two superfluid phases with smaller and larger condensates. The dotted dashed black line is extrapolated from the solid black line. The two dashed black lines are the turning points of the condensate curves with respect to $\rho$.}\label{Phasediagram_0.05}
\end{figure}

In Figure~\ref{Phasediagram_0.05}, we present the phase diagram of the system in the $\tau$-$\rho$ plane. The colors indicate the three different phases corresponding to the normal phase with zero condensate (gray), a first superfluid phase with small condensate (labelled SF phase $1$ and corresponding to light blue color) and a second superfluid phase with larger condensate (labelled SF phase $2$ and corresponding to light red color). The dashed red curve represents a line of second order critical points between the normal phase and the superfluid phase 1. This line is independent on the parameter $\tau$ \cite{Zhao:2022jvs}. The solid black line corresponds to a line of first order phase transition points which ends up at a critical point (CP) located at $(\rho_c=26.5784,\tau_c=0.000554)$. The dotdashed black line is a continuation of the line of first order phase transitions that, because of the low accuracy of the numerics in such region, is obtained using an extrapolation of the solid black curve. This dotdashed black line can also obtained using a thermodynamic landscape analysis (see details in Ref.~\cite{Zhao:2022jvs}). The two dashed black lines consist of the turning points in the condensate curves at different values of $\tau$ as a function of $\rho$, and they converge at the critical point. More precisely, the turning points are defined by the condition $\left(\partial \langle \mathcal{O}\rangle /\partial \rho\right)^{-1}=0$ (see for example Refs.~\cite{Zhao:2022jvs,Zhao:2023gcv}).

Above the critical point, in the region $\rho>\rho_c$ and $\tau>\tau_c$ there is no phase separation between the two superfluid phases depicted in blue and red colors. In other words, the two phases are continuously connected through the so-called ``supercritical region''. This situation is analogous to the supercritical phase in classical liquids in between the liquid and gas phases. One of the main task of this work is to investigate whether any separation line can be drawn in the supercritical region based on thermodynamic or dynamical arguments.
\section{Supercritical region and dynamical crossover}\label{dynsec}
We are interested in the supercritical region that appears in the phase diagram in the regime of large $\tau$ and large $\rho$. In particular, we want to question the existence of possible separation lines within the supercritical region depicted in Figure~\ref{Phasediagram_0.05}.

A first possibility, that was suggested long time ago by Frenkel in the context of classical liquids \cite{PhysRevLett.111.145901}, relies on the analysis of the late time dynamics and possible changes therein. In the context of classical liquids, this analysis is performed by studying the velocity auto-correlation function in time domain. Here, for this purpose, we study linear perturbations around these superfluid phases by calculating the quasi-normal mode (QNM) spectrum from the gravity side.

We notice that, in the context of holography, a similar dynamical transition was observed in the holographic superfluid model by changing temperature~\cite{Bhaseen:2012gg}. This dynamical transition is characterized by the change in the late time linearized dynamics, from a purely overdamped decay to an underdamped oscillating behavior. In the language of QNMs, overdamped decay corresponds to having the lowest non-hydrodynamic mode being purely imaginary. On the contrary, an oscillating underdamped behavior is the signal of the lowest non-hydrodynamic mode acquiring a finite real part, larger than its imaginary part (hence, underdamped). 

\begin{figure}[t]
	\center
\includegraphics[width=1\linewidth]{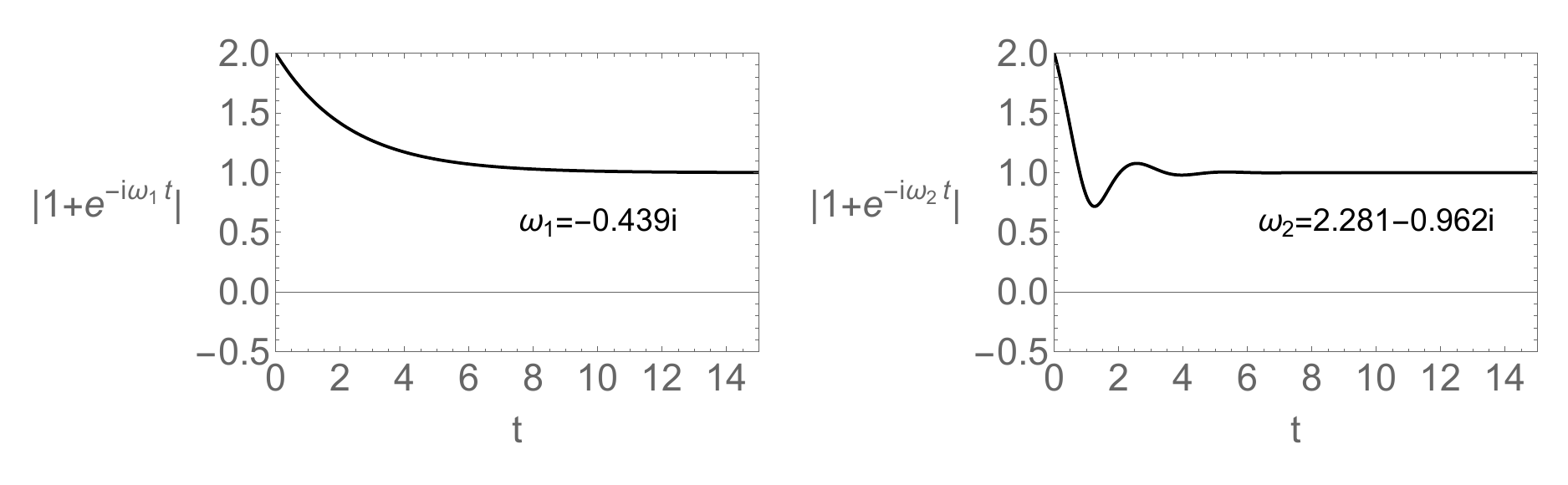}
	\caption{A schematic diagram of the difference between the dynamical behavior in the case of overdamped dynamics \textbf{(left)} and underdamped oscillating dynamics \textbf{(right)}. }\label{exp_display}
\end{figure}

The difference between these two behaviors is illustrated 
in Figure~\ref{exp_display}, using some fictitious frequency values. In the case of the purely imaginary mode, the time evolution $\propto \exp{-i \omega t}$ is a monotonically decreasing function of time, while in the case of the underdamped oscillating mode shows a non-monotonically evolution with attenuate oscillations. We notice that the Frenkel crossover normally discussed in classical liquids is slightly different as it corresponds to the appearance of a minimum in the time dependent velocity auto-correlation function. The presence of a minimum corresponds to an excitation having a finite real part but not necessarily much larger than its imaginary part.

As already anticipated, and explained in detail in Ref.~\cite{Bhaseen:2012gg}, the dynamical behavior at late time depends on the nature of the lowest non-hydrodynamic quasinormal mode in the complex $\omega$ plane. Assuming a separation of scales between the least damped mode and other non-hydrodynamic excitations, the late time dynamics is just given by $\text{exp}(-i\omega^* t)$, with $\omega^*$ the complex frequency of the lowest non-hydro mode. In the case of second order superfluid phase transition considered in Ref.~\cite{Bhaseen:2012gg}, close to the critical point the lowest non-hydrodynamic mode is the amplitude mode that displays a purely imaginary frequency \cite{Donos:2022xfd}.
\begin{figure}[ht]
	\center
\includegraphics[width=1\columnwidth]{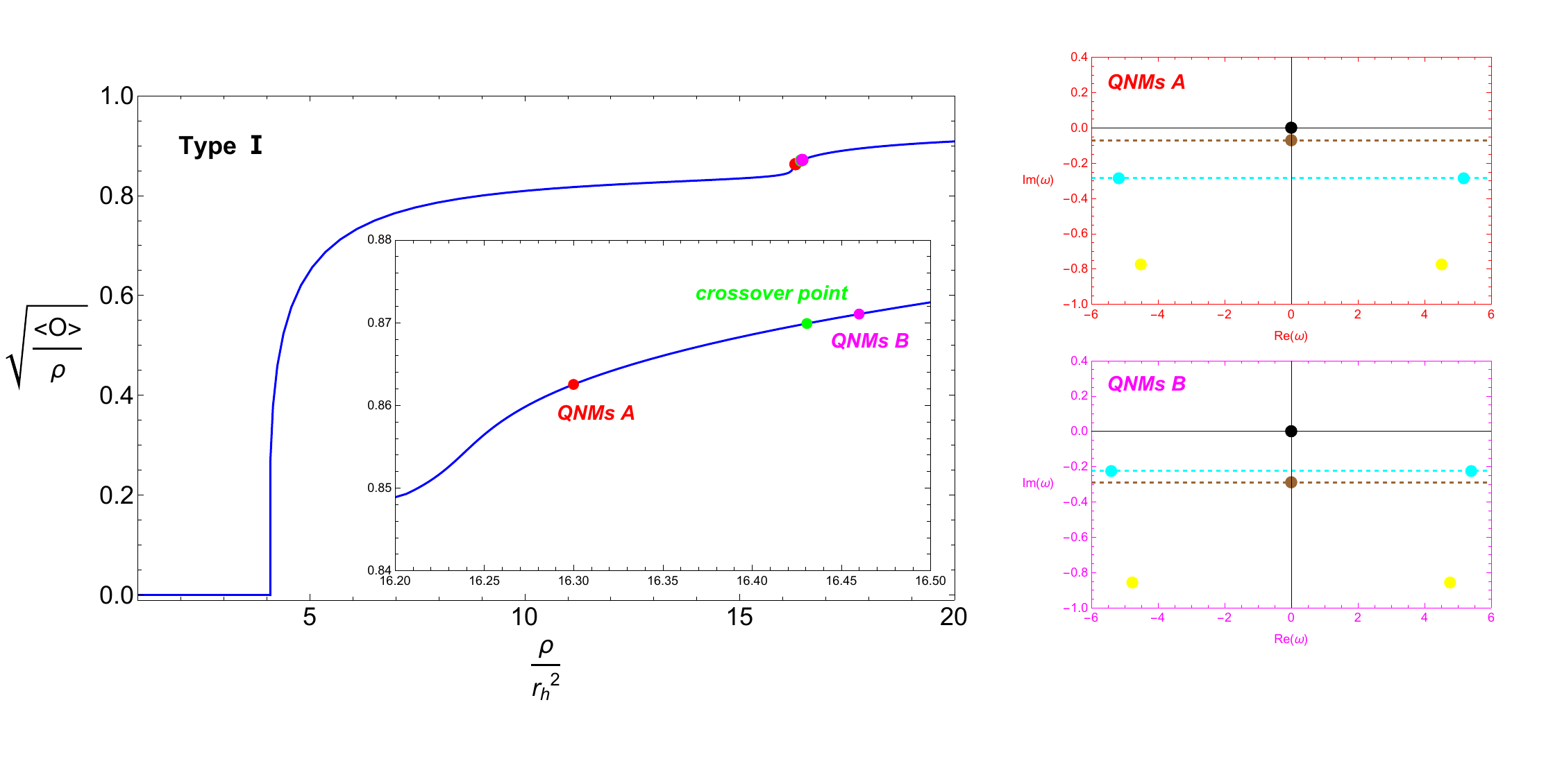}
	\caption{The condensate curve and QNMs for the Type I dynamical crossover.~\textbf{Left:} The order parameter as a function of the normalized charge density for $\lambda=-0.092$ and $\tau=0.00179$. Along the condensate line, red point: $\rho=16.3$ and $\sqrt{\langle \mathcal{O} \rangle /\rho}=0.862$; green point: $\rho=16.4307$, and $\sqrt{\langle \mathcal{O} \rangle /\rho}=0.870$; magenta point: $\rho=16.46$ and $\sqrt{\langle \mathcal{O} \rangle /\rho}=0.871$. The green dot corresponds to the location of the dynamical crossover. \textbf{Right:} The lowest quasinormal modes around the solutions depicted with red (A) and magenta (B) in the left panel. The black point is the Goldstone mode, the brown point is the amplitude mode, the yellow points and cyan points are the first and second pairs of higher non-hydro modes respectively.}\label{quasinormal_modes_imaginary_plane}
\end{figure}

Moving away from the critical point into the region with larger condensate, the amplitude mode is moving downwards along the imaginary axes, while a couple of non-hydrodynamic modes with opposite real part of the frequency $\text{Re}(\omega)$ go upwards from below. Then at some point in the superfluid phase with finite condensate, the imaginary part of these two types of modes become equal, signaling the onset of the dynamical crossover. The change in the lowest non-hydrodynamic modes results in the change of the late time behavior as displayed in Figure~\ref{exp_display}. 

A recent study~\cite{Zhao:2022jvs} based on the action in Eq.~(\ref{Lagm}) allows for a more intricate phase diagram and for several phase transitions including first order phase transitions between the normal phase and superfluid phase, as well as a ``cave of wind'' phase transition which is a first order phase transition between two superfluid phases with higher and lower condensates. In this setup, the QNMs have been studied to characterize the linear stability of the system in the canonical ensemble and have been later verified by performing the fully nonlinear dynamical evolution~\cite{Zhao:2023gcv}.
\begin{figure}[ht]
	\center
\includegraphics[width=1\columnwidth]{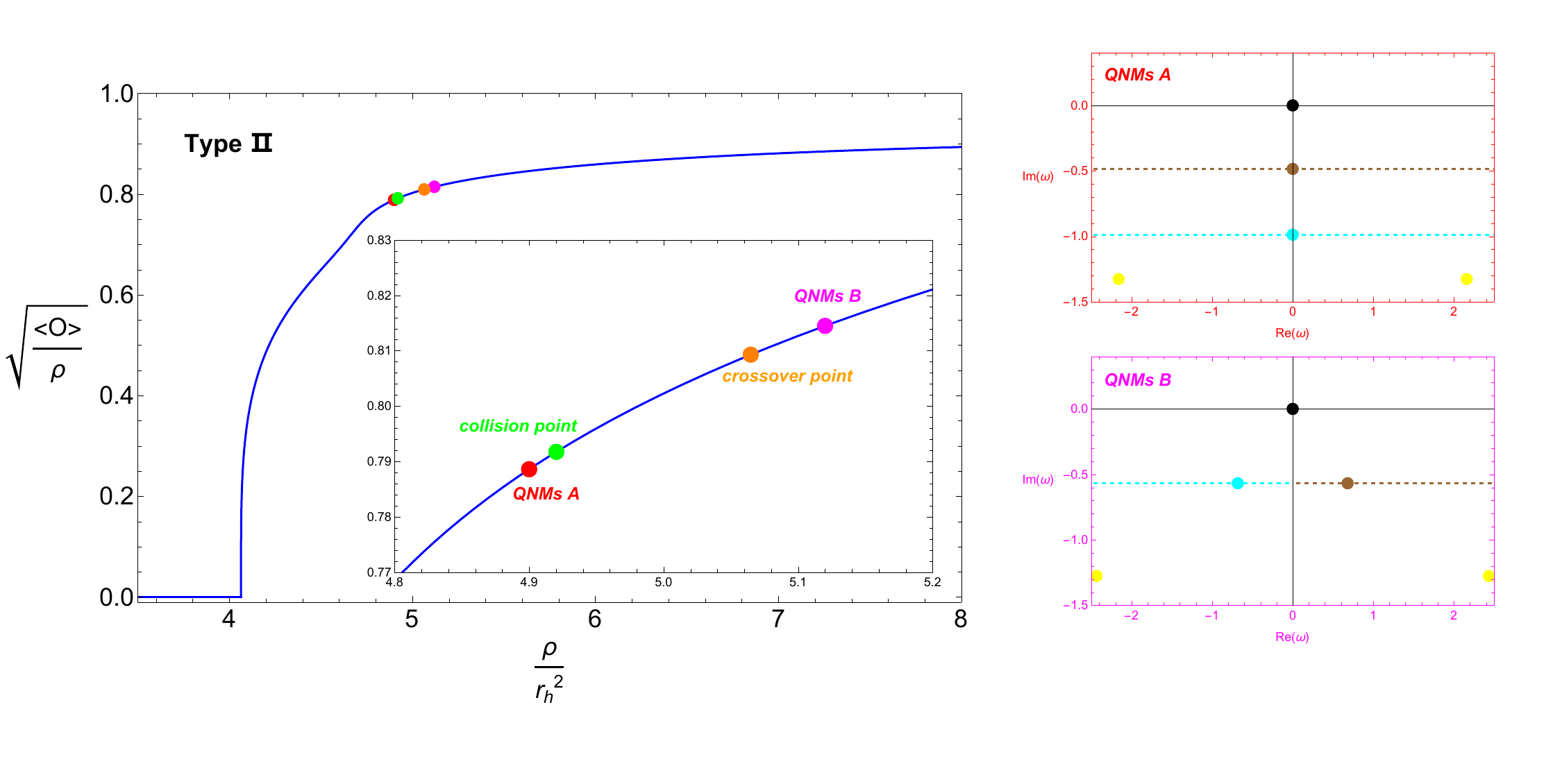}
	\caption{The condensate curve and QNMs for the Type II dynamical crossover.~\textbf{Left:} The order parameter as a function of the normalized charge density for $\lambda=-0.8$ and $\tau=0.18$. Red point: $\rho=4.90$ and $\sqrt{\langle \mathcal{O} \rangle /\rho}=0.789$. Green point: $\rho=4.92$ and $\sqrt{\langle \mathcal{O} \rangle /\rho}=0.792$. Orange point: $\rho=5.06473$ and $\sqrt{\langle \mathcal{O} \rangle /\rho}=0.809299$. Magenta point: $\rho=5.12$ and $\sqrt{\langle \mathcal{O} \rangle /\rho}=0.814501$. \textbf{Right:} The corresponding quasinormal modes around solution A (red) and solution B (magenta). The collision between the amplitude mode (brown) and a second purely imaginary mode (light blue) is evident. This collision happens at the point marked by the green dot in the left panel. However, the dynamical crossover happens in a different location defined by the requirement $|\text{Re}(\omega)|=|\text{Im}(\omega)|$ and indicated with an orange symbol.}\label{quasinormal_modes_plane_Typetwo}
\end{figure}

In this model, two different types of dynamical crossover can be defined. A first type I dynamical crossover is analogous to what observed in Ref.~\cite{Bhaseen:2012gg} and it is defined by the transition between a purely imaginary pole and a couple of non-hydrodynamic complex poles. An example of type I dynamical transition is shown in Figure~\ref{quasinormal_modes_imaginary_plane} between two concrete points corresponding to the red and magenta dots on the condensate curve. As evident from the right panel, around solution A (red point) the amplitude mode is the lowest non-hydro mode. Nevertheless, by increasing further the charge density and moving to solution B (magenta point), the first pair of non-hydro modes (indicated with yellow dots) moves upwards and at a critical value they become the lowest dominant non-hydro modes. This transition happens exactly at the crossover point that is depicted in the left panel in green color.

A different type of dynamical transition can also take place. We label it type II. This dynamical crossover is defined by the collision between the amplitude mode and another purely imaginary mode. As a consequence of this collision, the two imaginary modes transform into a couple of oscillating modes with finite and opposite values of real frequency $\text{Re}(\omega)$. When the real part of these modes become larger than their imaginary part, the corresponding late-time dynamics become oscillating rather than monotonically decaying, signaling the onset of the dynamical crossover.
An example of this second dynamical crossover is shown in Figure~\ref{quasinormal_modes_plane_Typetwo}. As we will see later, in our system the type I dynamical transition always anticipates the type II.

The dynamical crossover is usually considered near a critical point corresponding to a second order phase transition (see \textit{e.g.}, Ref.~\cite{Bhaseen:2012gg}) and it is closely related to the presence of a overdamped imaginary mode near the critical point that is then superseded by a couple of modes with finite real part moving away from it (see Figure~\ref{quasinormal_modes_imaginary_plane}). Interestingly, recent results for first order superfluid phase transitions~\cite{Zhao:2022jvs} in the model considered in this manuscript show that the imaginary part of the amplitude mode always goes to zero when the solution approaches the turning points of the condensate curve, where $\left(\partial \langle \mathcal{O}\rangle /\partial \rho\right)^{-1}=0$. In this sense, these turning points share with second order critical points the property of having a overdamped imaginary mode whose damping coefficient grows moving away from them, where, on the other hand, it vanishes. Therefore, it is natural to expect that a similar dynamical crossover might be also observed near these turning points in first order phase transition scenarios. 

\begin{figure}[ht]
	\center
\includegraphics[width=1\columnwidth]{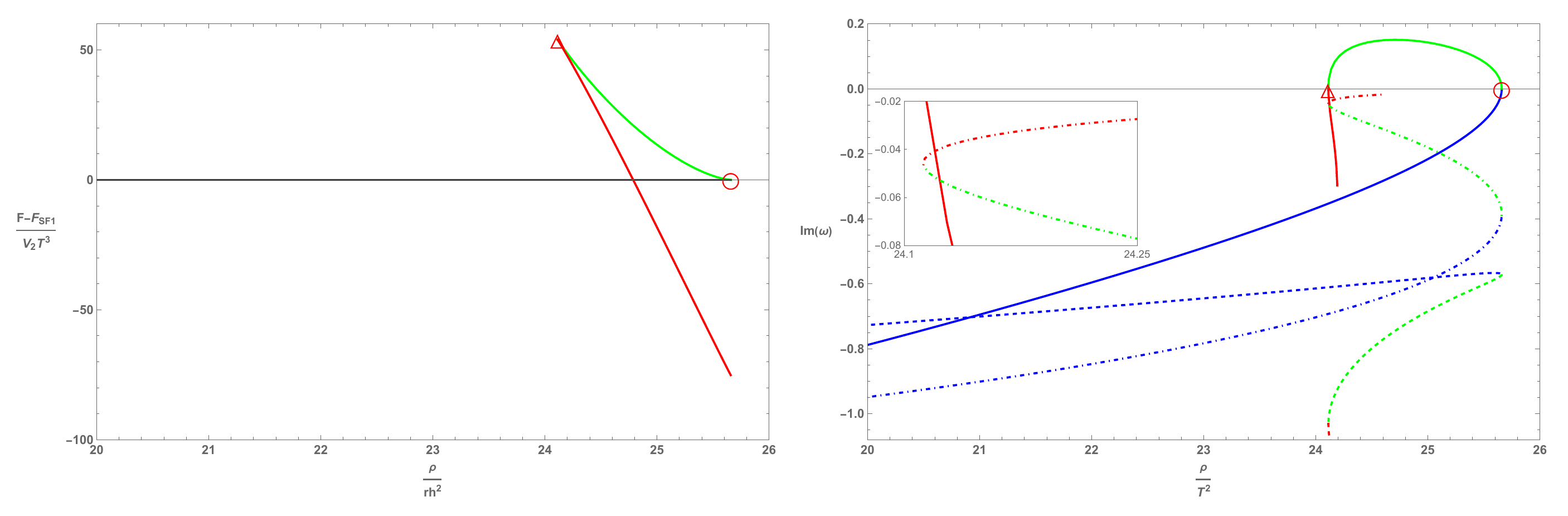}
\caption{The relative value of free energy as well as the imaginary part of the QNMs in a first order phase transition.~\textbf{Left:} The relative value of free energy with respect to the SF1 solution with lower condensate (blue line) as a function of normalized charge density for $\lambda=-0.05$ and $\tau=0.0005$.~\textbf{Right:} imaginary part of the lowest QNMs involved in the dynamical crossovers as a function of normalized charge density.\label{imomega_lambda0.05}}
\end{figure}

Because of these similar dynamical properties, we define these turning points in the condensate curve as ``quasi-critical points''. These quasi-critical points are also located at the two tips of the swallow tail shaped free energy curve (see left panel in Figure~\ref{imomega_lambda0.05}), that is a universal feature in first order phase transitions.

The phase transition points in first order phase transitions are defined as the places where the two branches of stable (meta-stable) solutions have the same free energy. On the other hand, the two quasi-critical points are each connected to one branch of solutions, which is meta-stable before the phase transition point and stable after. Right at the quasi-critical points, the imaginary part of the lowest QNM is zero, $\text{Im}(\omega)=0$. Therefore, away from one of the quasi-critical points along the meta-stable to stable branch of solutions, this purely imaginary mode is moving down along the imaginary axes away from the origin, a behavior that may result in a dynamical crossover of either type I or type II as described above. As a concrete scenario, let us consider the case with ($\lambda=-0.05,~\tau=0.0005$).

In the left panel of Figure~\ref{imomega_lambda0.05}, we show the free energy versus the charge density for the superfluid solutions near the first order phase transition point, which show a standard swallow tail type. The red and blue curves denote the two branches of stable (meta-stable) solutions, while the green curve denotes the branch of unstable solutions. In the right panel of the same figure, we show the value of $\text{Im}(\omega)$ for the overdamped amplitude mode, as well as for the underdamped oscillating modes involved in the dynamical crossover, on the two branches of stable (meta-stable) solutions. We use solid lines to denote the pure imaginary amplitude mode which crosses the horizontal axes at the  quasi-critical points, and use dashed lines to show the imaginary part of the pairs of oscillating modes with finite real part as well. The color scheme represents the different solutions whose free energy is shown in the left panel of the same figure. 

This analysis confirms that the imaginary part of the amplitude mode vanishes at the two quasi-critical points located at the tips of the red and blue branches, respectively. Away from the quasi-critical point at higher charge density denoted by the red circle, and along the blue branch branch of solutions, the amplitude mode moves downwards along the imaginary axes as shown by the solid blue line. On the other hand, the dashed blue lines show the imaginary part of the pair of underdamped oscillating modes with finite real part. We can see that the solid and dashed blue lines intersect at one specific point $\rho=\rho_D$, which indicates the onset of a type I dynamical crossover. A similar situation appears for the red branch of solutions connected to the quasi-critical point at lower charge density marked by the red triangle. We notice that, in the red case, the dynamical crossover happens very close to the quasi-critical point. In order to visualize this better we provide a zoom of this crossover in the inset of the right panel in Figure~\ref{imomega_lambda0.05}.
\begin{figure}[t]
	\center
\includegraphics[width=0.47\columnwidth]{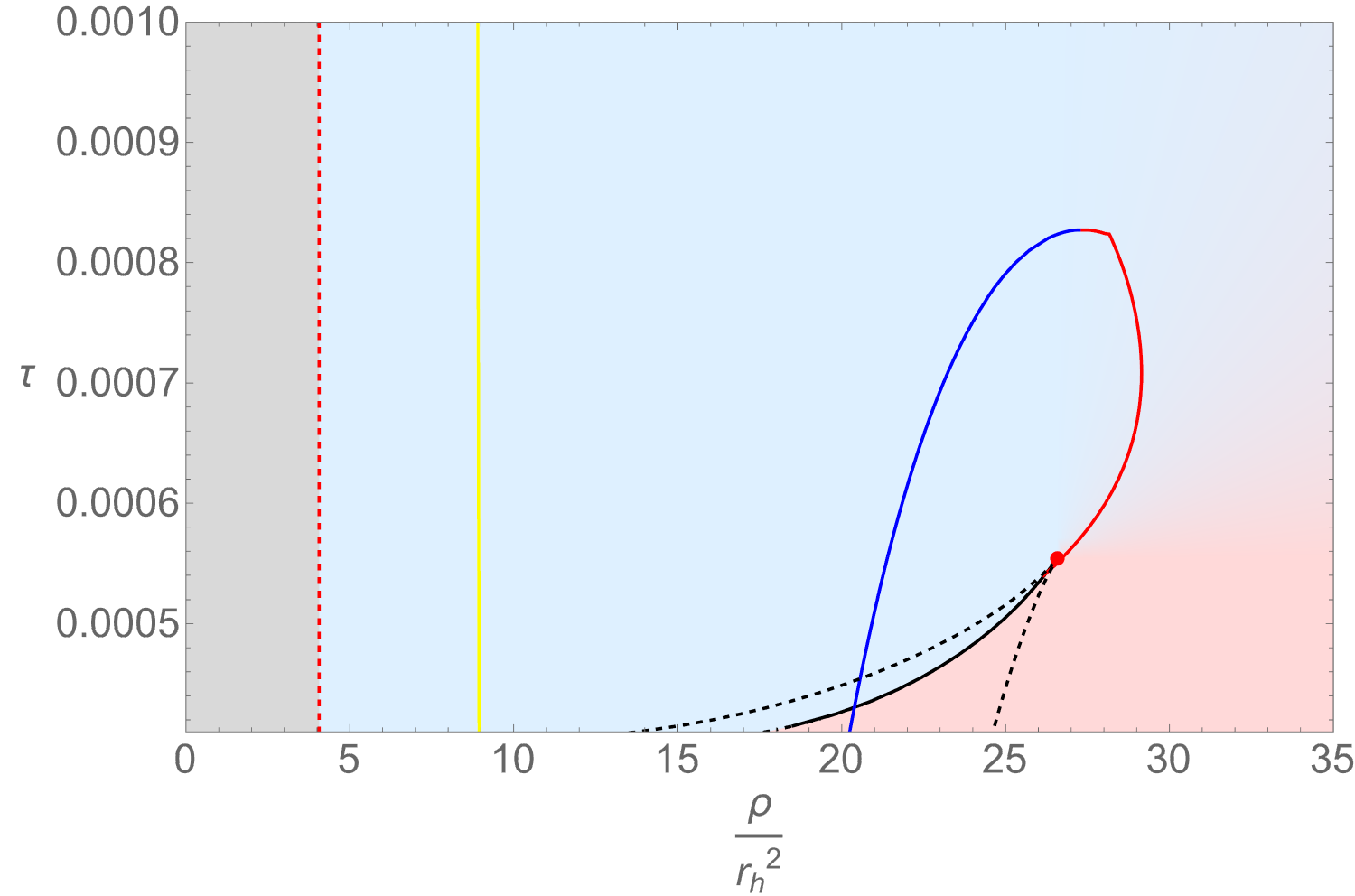}
\includegraphics[width=0.47\columnwidth]{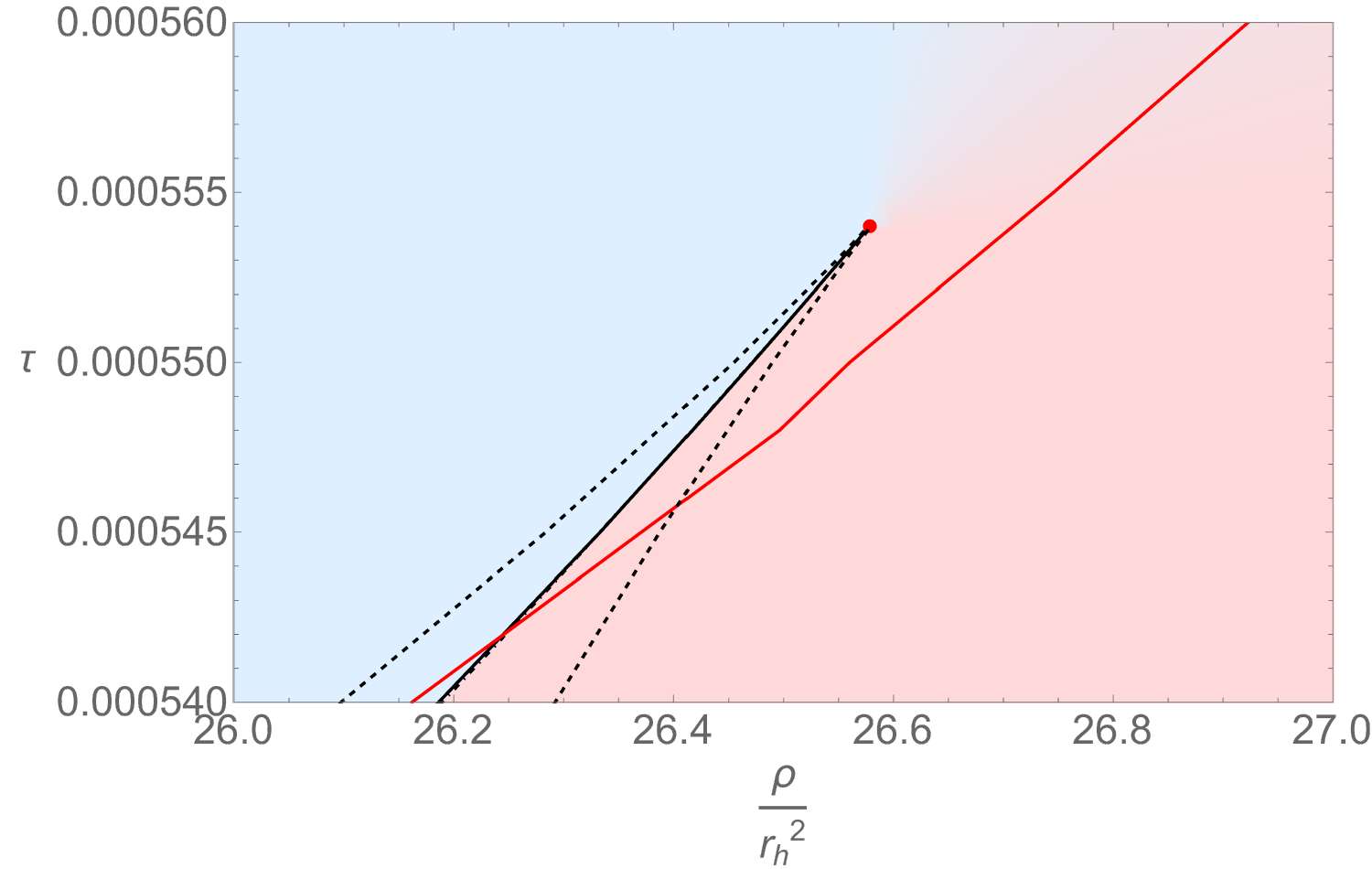}
	\caption{The revised phase diagram for $\lambda=-0.05$ to be compared with the original one in Figure~\ref{Phasediagram_0.05}. The yellow, red and blue lines indicate the dynamical crossover points as discussed in the main text. The left panel presents a zoom of the region around the supercritical point.}\label{lambda0.05}
\end{figure}
\section{A revised phase diagram}\label{sec4}

After analyzing in detail the dynamical crossovers near the critical and quasi-critical points, we are now in the position to add more information on the phase diagram.

Let us consider the case of $\lambda=-0.05$ as an concrete example. The bare phase diagram has been already presented in Figure~\ref{Phasediagram_0.05}, where the critical point $(\rho_c=26.5784,\tau_c=0.000554)$ is indicated with a red symbol. In Figure~\ref{lambda0.05}, we revisit the same phase diagram by adding on top of it the dynamical crossover lines defined and discussed in the previous section.

In the left panel of Figure~\ref{lambda0.05}, the solid red and blue lines indicate the dynamical crossover points in the vicinity of the critical point. The union of the blue and red lines form a closed region in the phase diagram where the late time dynamics are dominated by the overdamped mode, and decay monotonically following an exponential form. On the contrary, in the region outside the solid red and blue lines the late dynamics are oscillatory since they are governed by a pair of underdamped modes with finite real part. In the right panel of Figure~\ref{lambda0.05}, we provide an enlarged view around the critical point to prove that the red ``Frenkel line'' does not terminate on the critical point. This is analogous to what happens in realistic classical liquids, where the dynamical crossover lines do not end at the critical point (see \textit{e.g.}, Ref.~\cite{li2023thermodynamic}). Furthermore, we see that there are two lines of dynamical crossover (``Frenkel lines''), indicated respectively with blue and red color. This is different from the situation in classical supercritical fluids. The two lines here appear because both phases (light blue and light red regions) are superfluid states that far from the critical point display oscillating modes. This, on the contrary, it is not true deep in the gas phase of classical supercritical fluids where the dynamics are strongly overdamped and the velocity auto-correlation function decays monotonically.
\begin{figure}[ht]
	\center
\includegraphics[width=0.47\columnwidth]{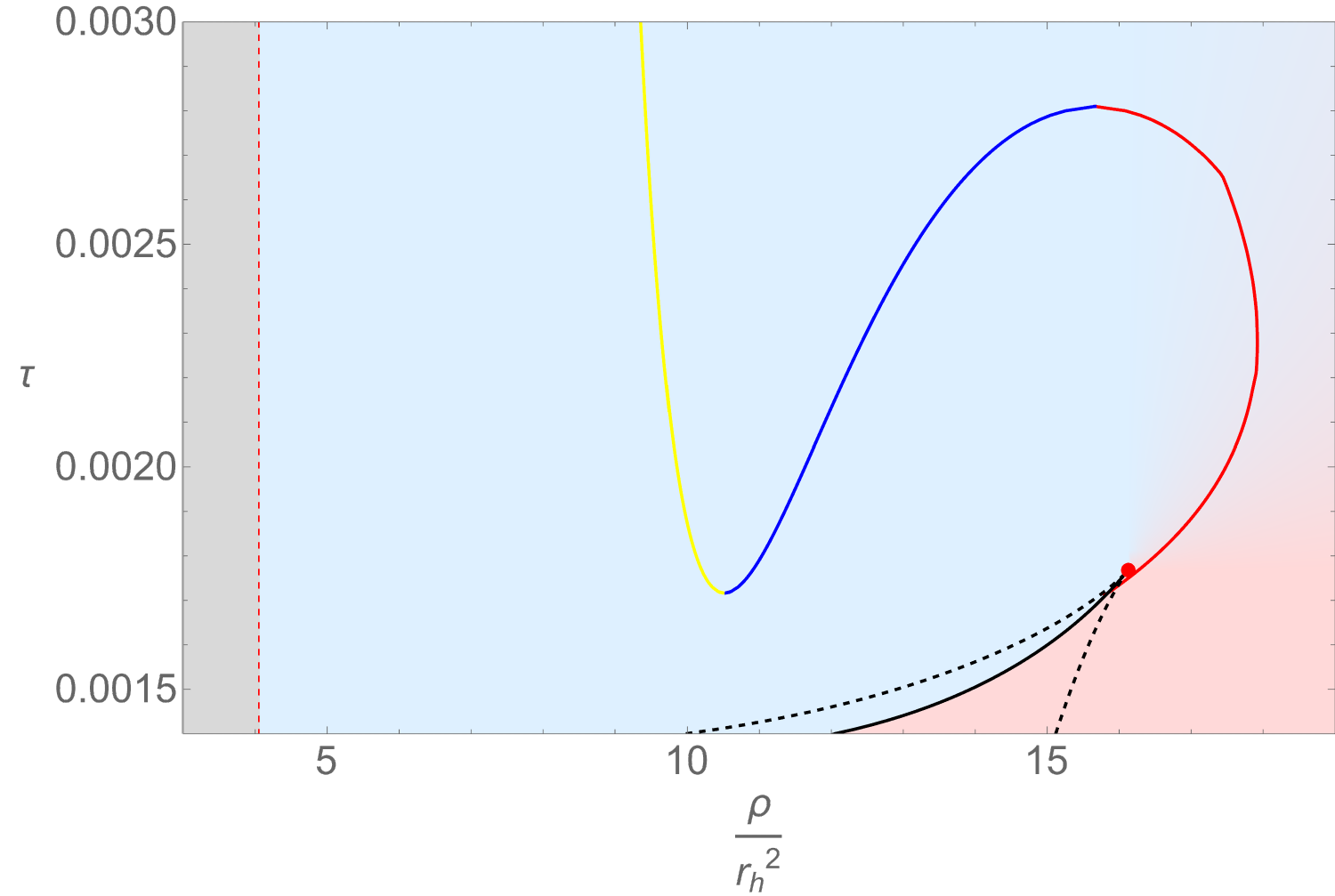}
\includegraphics[width=0.47\columnwidth]{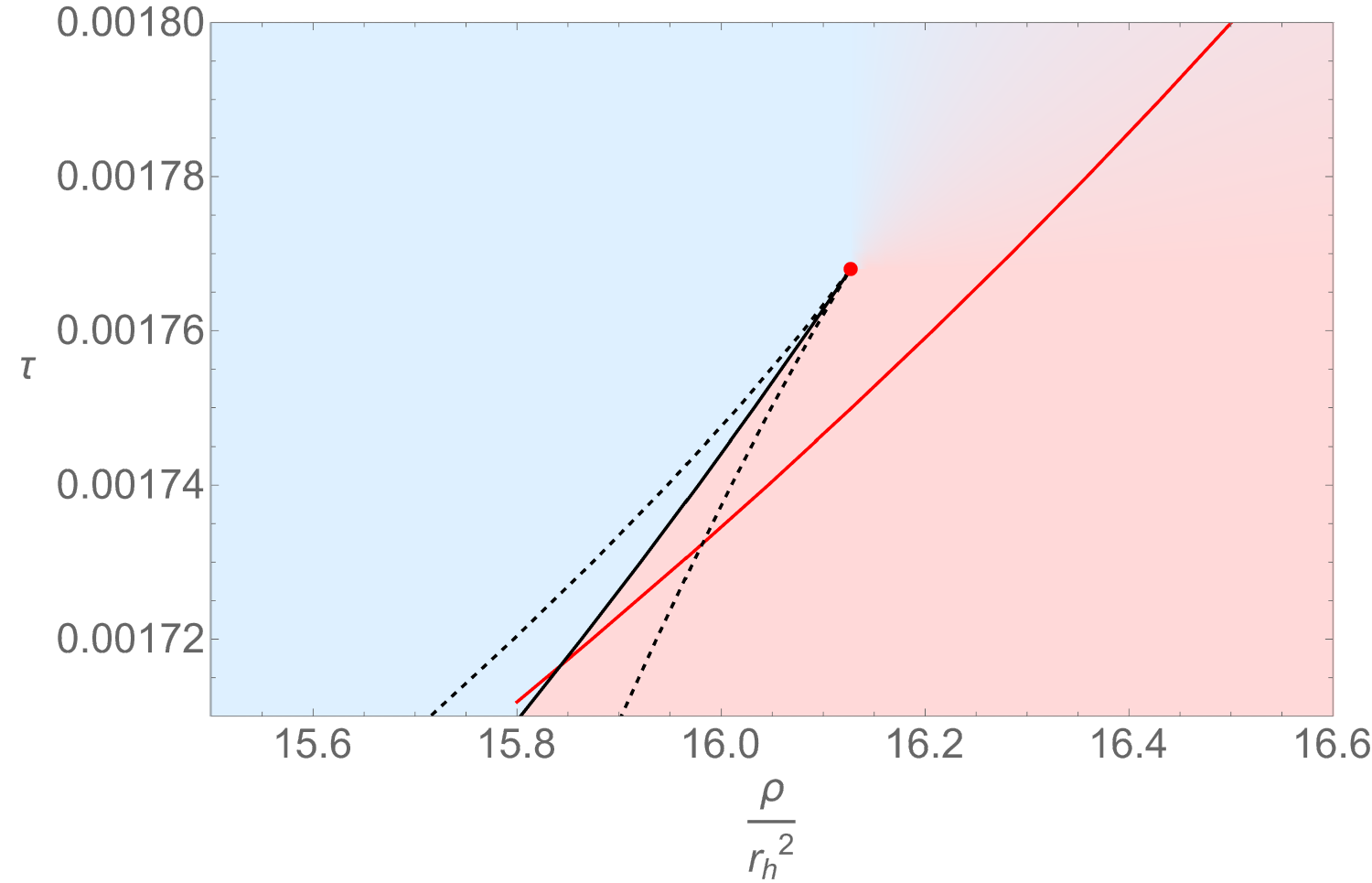}
	\caption{The phase diagram for $\lambda=-0.092$. The meaning of the various lines is the same as in Figures~\ref{Phasediagram_0.05} and \ref{lambda0.05}.}\label{critical_point_0.092}
\end{figure}

In the left panel of Figure~\ref{lambda0.05}, we can also observe a vertical solid yellow line, indicating the presence of a secondary dynamical crossover. This transition is a consequence of the line of second order critical points that is visible at lower charge density and indicated with a dashed red line. Approaching the second order superfluid transition, the dynamics are once again dominated by a purely imaginary mode whose imaginary part becomes smaller and smaller moving towards the red dashed line. hence, moving towards small charge density it is inevitable that at after a critical value the dynamics become dominated by a purely imaginary mode again.

The structure of the dynamical crossover lines on the phase diagram depends crucially on the potential parameter $\lambda$. For example, in some cases, the yellow and blue crossovers lines in Figure~\ref{lambda0.05} can join together. As an example of this special case, we plot the $\tau-\rho$ phase diagram for $\lambda=-0.092$ in Figure~\ref{critical_point_0.092}. In this case, there is a big region, below the yellow and blue curves in which the dynamics are dominated by an overdamped pure imaginary mode.

Finally, we notice that in both the phase diagrams in Figure~\ref{lambda0.05} and Figure~\ref{critical_point_0.092}, the dynamical crossovers are all of type I.
\section{Thermodynamic signatures in the supercritical region}\label{termsec}
After discussing in detail the existence and role of dynamical separation lines within the supercritical region of our superfluid system, we are now interested in determining whether any thermodynamic criterion can be used to separate the two phases as well. In doing so, we take inspiration from the idea of Widom line \cite{doi:10.1073/pnas.0507870102} and symmetry line \cite{doi:10.1021/acs.jpcb.9b04058} introduced in the context of classical supercritical fluids, but also on the idea developed in nuclear matter of using scaled variance, skewness, and kurtosis as probes for the QCD critical point \cite{PhysRevLett.102.032301,PhysRevLett.107.052301,PhysRevC.92.054901}.

In general, the derivatives of the free energy reflect important physical properties of critical points, and most importantly their order. For instance, in a first order phase transition, the first derivative at the transition point is discontinuous, the second derivative diverges, and the third derivative can differentiate between different phases based on its sign. Higher order derivatives of the free energy in the supercritical region might serve a similar purpose, as already discussed in the past \cite{doi:10.1021/acs.jpcb.9b04058,PhysRevLett.102.032301,PhysRevLett.107.052301,PhysRevC.92.054901}. For this purpose, we define with $k_n$ the n-th derivative of the free energy with respect to the charge density $\rho$,
\begin{align}
k_n \equiv \frac{\partial^n F(\rho)}{\partial \rho ^n}~.
\end{align}
Here, $k_2$ is the variance, $k_3$ the skewness, $k_4$ the kurtosis and so on.

\begin{figure}[t]
	\center
\includegraphics[width=\columnwidth]{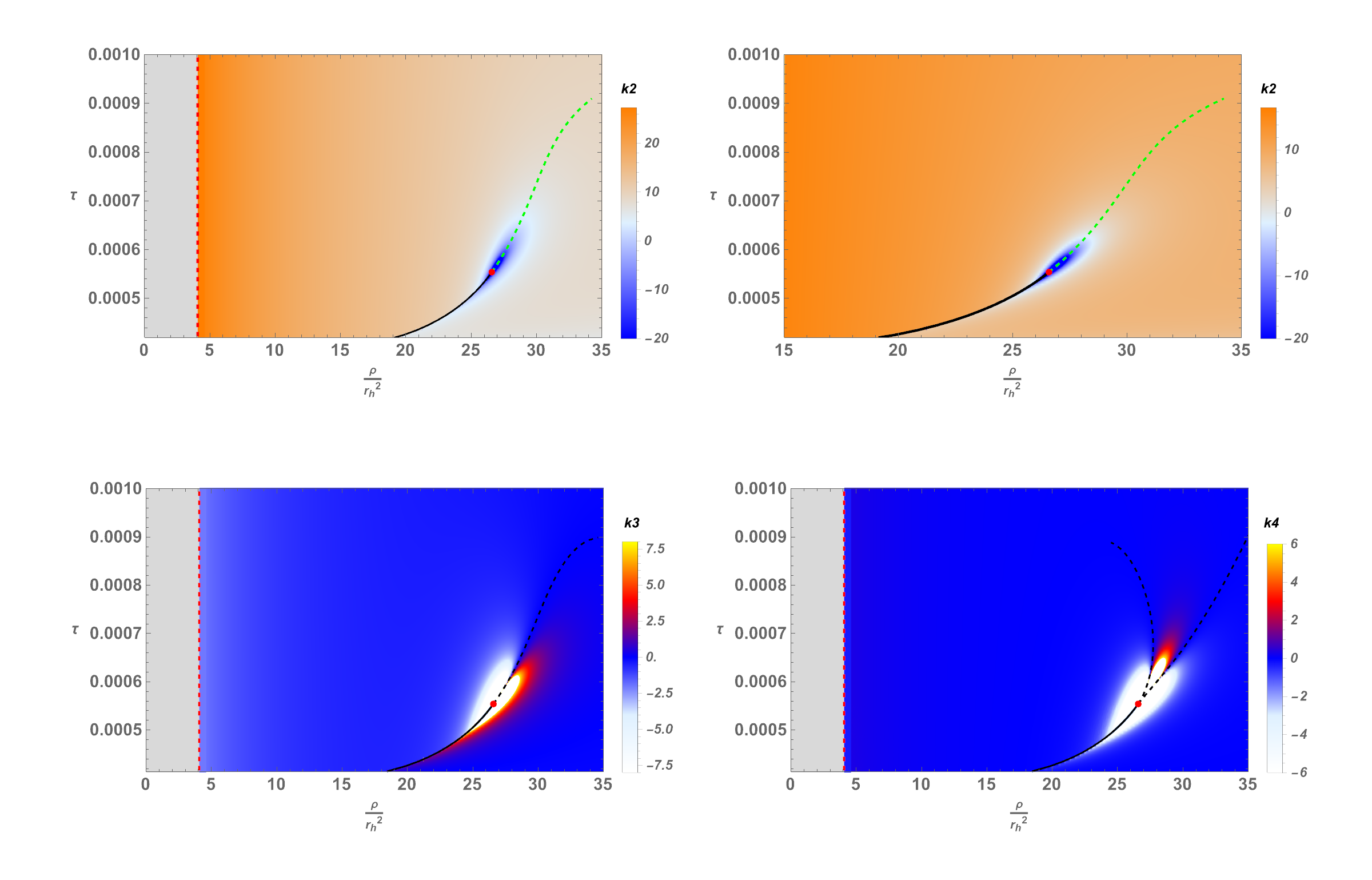}
	\caption{Thermodynamic signatures in the supercritical region for $\lambda=-0.05$. The gray region is the normal phase that is separated from the superfluid phase by a line of second order critical points (red dashed). The black solid line is the line of first order phase transition points. The red point is the supercritical point at the end of the first order line. The meaning of the color maps are indicated with bar legends. In the bottom panels, the black dashed line represents points where the corresponding thermodynamic function vanishes. In the top panels the dashed green lines indicate the location of the minima of $k_2$ at fixed values of $\tau$.}\label{freeOMEGA0o05}
\end{figure}

We have analyzed these thermodynamic quantities in the whole phase diagram for the case of $\lambda=-0.05$ in Figure~\ref{freeOMEGA0o05} (we have explicitly checked that other choices of $\lambda$ lead to similar results). In the top panels, we show the value of the variance in the whole phase diagram. At exactly the critical point, indicated with a red symbol, $k_2$ diverges as one might expect for a second order critical point. In the vicinity of the critical point, we observe that the variance displays a maximum that moves continuously away from the critical point as indicated by the blue region. On the right and left side of this blue region the value $k_2$ increases. This implies that close to the critical point, even if $k_2$ does not diverge anymore, it still displays a minimum that allows one to draw a separation line between the two phases emanating from the critical point. The line of minima is indicated with green dashed lines in the top panels of Figure~\ref{freeOMEGA0o05}. In other words, following the location of the minimum of $k_2$, we can identify a separation line between the two phases. Far away from the critical point, the minima disappear and this separation line cannot be clearly identified anymore. Our findings are analogous to the results in nuclear matter presented in Ref.~\cite{PhysRevC.92.054901} and relate to the definition of the Widom line in classical supercritical fluids \cite{doi:10.1073/pnas.0507870102}. Finally, we notice that $k_2$ becomes also very large by approaching the other line of second order critical points indicated by the red dashed line in the top left panel in Figure~\ref{freeOMEGA0o05}. This is natural since on this line, that separates the normal phase (gray region) from the superfluid phase (orange color), $k_2$ has to diverge.

In the bottom panels of the same figure, we display the behavior of the skewness and kurtosis parameters focusing on the region near the supercritical point. Similarly to the scaled variance, both these parameters diverge at the supercritical point but they present a much richer structure that has already been very useful in the study of supercritical systems (see \textit{e.g.}, Refs.~\cite{doi:10.1021/acs.jpcb.9b04058,PhysRevLett.102.032301,PhysRevLett.107.052301,PhysRevC.92.054901}). The line on which the skewness parameter $k_3$ is zero (indicated with dashed black line) emanates from the supercritical point and divides the supercritical region in two separate regions. Importantly, the sign of the skewness is opposite on the two sides of this separation line and it is negative in the phase akin to the liquid phase and positive in the gas-like phase. This dipolar structure is very strong in the region close to the supercritical point. This result is consistent with what found in the QCD phase diagram \cite{PhysRevC.92.054901}. Finally, we discuss the kurtosis parameter $k_4$ in the bottom right panel of Figure~\ref{freeOMEGA0o05}. We observe that the zeros of this thermodynamic parameter define two different separation lines that emanate from the critical point and extend into the supercritical region (see black dashed lines). These lines form an intermediate crossover region between the two phases where the kurtosis is strongly negative. This feature was proposed as a possible probe for the QCD critical point in the past~\cite{PhysRevLett.107.052301}. Also, the existence of two different thermodynamic separation lines is consistent with the expectations from the Ising model, as recently advocated in Ref.~\cite{li2023thermodynamic}. Moreover, in the two regions outside these separation lines, and close to the supercritical point, the kurtosis is very large and positive.

We conclude with some observations. First, all the thermodynamic separation lines emanate from the critical point. This is remarkably different from the dynamical crossovers discussed in the previous section. Second, the dynamical and thermodynamic criteria do not agree with each other and provide different separation lines withing the supercritical region. Finally, the thermodynamic criteria, and in particular the behavior of the kurtosis, introduce two different separation lines and an intermediate crossover region between the liquid-like and gas-like phases.
\section{Conclusion}\label{sec5}
In this work, we have considered a holographic superfluid model with non-linear bulk interactions that presents a rich phase diagram including second order and first order critical points. In the regime of large charge density, the phase diagram of this model is very similar to that of standard fluids as it displays a line of first order phase transition point that terminates on a critical point above which the so-called supercritical region extends. Qualitatively, this scenario is analogous to the separation between the liquid phase and the gas phase in molecular fluids. It also shares several similarities with the predicted phase diagram of QCD and its critical point (see Figure~\ref{fig:0}).

Within the supercritical region of this system, we have explored the existence of dynamical separation lines based on a sudden change of dynamics not accompanied by any thermodynamic feature. This type of dynamical crossover is known in the literature as Frenkel line \cite{PhysRevLett.111.145901}. By studying the spectrum of quasinormal modes, we have been able to connect the concept of Frenkel line to the behavior of the low-energy excitations and in particular to a crossover between a lowest lying purely imaginary QNM and a pair of modes with finite real part, that has been already observed in different circumstances in Ref.~\cite{Bhaseen:2012gg}. In our study, we generally observe two Frenkel lines on the two sides of the critical point. This is different from the classical liquid-gas transition and it happens because both phases on the left and right of the critical point are superfluid states where oscillating modes can emerge. On the contrary, this is not the case for gases where the dynamics are overdamped and the velocity auto-correlation function decays monotonically.

Finally, we have studied higher-order thermodynamic susceptibilities in order to find possible thermodynamic lines separating the two phases withing the supercritical region. In agreement with previous analyses in classical liquids \cite{doi:10.1021/acs.jpcb.9b04058} and nuclear matter \cite{PhysRevLett.102.032301,PhysRevLett.107.052301,PhysRevC.92.054901}, these higher momenta allow to identify well-defined thermodynamic separation lines. Interestingly, one of them (kurtosis) predicts the existence of two different separation lines and of an intermediate region between the two phases. This is similar to what recently advocated in classical liquids by using the analogy with the Ising model phase diagram \cite{li2023thermodynamic}. Finally, we notice that higher-order thermodynamic susceptibilites have been already considered in the context of holographic QCD \cite{Fu:2024wkn,Zhao:2023gur}.

Our results suggest that the presence of dynamical and thermodynamic crossover lines is an ubiquitous feature of systems displaying a complex phase diagram with a supercritical region. It would be interesting to understand further the consequences of these separation lines on different transport properties such as the viscosity, and other dynamical observables such as the speed of sound that can be experimentally measured (see \textit{e.g.}, Refs.~\cite{jiang2024experimental,2023PhRvR...5a3149H,PhysRevE.107.055211}). We leave these questions for the near future.
\section*{Acknowledgements}
MB would like to thank Yuliang Jin, Xinyang Li, Shi Pu, Song He, Li Li and Sha Jin for valuable discussions.
This work is partially supported by the National Natural Science Foundation of China (Grant Nos.11965013, 11975072, 11875102, and 11835009), the National SKA Program of China (Grants
Nos. 2022SKA0110200 and 2022SKA0110203)
and the National 111 Project (Grant No. B16009).
ZYN is partially supported by Yunnan High-level Talent Training Support Plan Young $\&$ Elite Talents Project (Grant No. YNWR-QNBJ-2018-181). MB acknowledges the support of the Shanghai Municipal Science and Technology Major Project (Grant No.2019SHZDZX01) and of the sponsorship from the Yangyang Development Fund.
\bibliographystyle{JHEP}
\bibliography{reference}
\end{document}